# Van der Waals Epitaxy of Two-Dimensional Single-Layer h-BN on Graphite by Molecular Beam Epitaxy: Electronic Properties and Band Structure

Debora Pierucci[1], Jihene Zribi[2], Hugo Henck[2], Julien Chaste[2], Mathieu G. Silly[3], François Bertran[3], Patrick Le Fevre[3], Bernard Gil[4,5], Alex Summerfield[6], Peter H. Beton[6], Sergei V. Novikov[6], Guillaume Cassabois[4], Julien E. Rault[3], and Abdelkarim Ouerghi[2]

[1]CELLS - ALBA Synchrotron Radiation Facility, Carrer de la Llum 2-26, 08290 Cerdanyola del Valles, Barcelona, Spain
[2]Centre de Nanosciences et de Nanotechnologies, CNRS, Univ. Paris-Sud, Université Paris-Saclay, C2N – Marcoussis, 91460 Marcoussis, France
[3]Synchrotron-SOLEIL, Saint-Aubin, BP48, F91192 Gif sur Yvette Cedex, France
[4] Laboratoire Charles Coulomb UMR 5221, Université de Montpellier & CNRS, F-34095, Montpellier, France
*5- Ioffe Institute, St. Petersburg, 194021, Russia*
[6]School of Physics and Astronomy, University of Nottingham, Nottingham NG7 2RD, United Kingdom

We report on the controlled growth of h-BN/graphite by means of molecular beam epitaxy (MBE). X-ray photoelectron spectroscopy (XPS) suggests an interface without any reaction or intermixing, while the angle resolved photoemission spectroscopy (ARPES) measurements show that the h-BN layers are epitaxially aligned with graphite. A well-defined band structure is revealed by ARPES measurement, reflecting the high quality of the h-BN films. The measured valence band maximum (VBM) located at 2.8 eV below the Fermi level reveals the presence of undoped h-BN films (band gap ~ 6 eV). These results demonstrate that, although only weak van der Waals interactions are present between h-BN and graphite, a long range ordering of h-BN can be obtained even on polycrystalline graphite *via* van der Waals epitaxy, offering the prospect of large area, single layer h-BN.

Boron nitride (BN) is a synthetic material that has attracted considerable interest over the past decade[1]. BN presents four polymorphs, arising from different bonding configurations of boron (*B*) and nitrogen (*N*) atoms: cubic (c-BN), hexagonal (h-BN), rhombohedral (r-BN) and wurtzite (w-BN), and the variety of properties of BN materials are strongly related to these crystal structures. The most stable crystalline form, hexagonal h-BN, is an insulating isomorph of graphite[2]. The boron and nitrogen atoms are arranged in two inequivalent A and B sublattices of a planar honeycomb structure formed by $sp^2$ covalent bonds. As for graphite, the bonding forces between the atomic planes are of van de Waals (vdW) type. However, the band structures of h-BN and graphite show very significant differences; the different onsite energies of the *B* and *N* atoms lead to a large (~ 6 eV) band gap and a small (1.8 %) lattice mismatch with graphite[3–6]. h-BN possesses many highly desirable physical and chemical properties such as low density, high melting point, high-temperature stability, high thermal conductivity, low dielectric constant and chemical inertness[7]. All these properties make h-BN a very interesting and unique material for various electronic and optoelectronic devices[8,9], either for use as a substrate or an insulating dielectric[10]. With the advent of graphene and two-dimensional (2D) materials in recent years, h-BN, as the only 2D-insulator[11,12], is an elementary building block which may be combined with other 2D materials to form complex vdW heterostructures, thus opening a new paradigm in the physics of 2D solids[13]. Moreover its structural similarity to graphene and its optical properties can be utilised to improve 2D devices[14,15]. To achieve the best possible performance of these new 2D material vdW heterostructures, h-BN with high crystalline quality is needed[16]. In semiconductor physics, it is well known that epitaxy is the most appropriate technique to obtain large area, high quality devices. However, in conventional semiconductor materials the requirement for lattice matching limits the materials that can be combined and the quality of interfaces that can be achieved. The use of 2D materials potentially circumvents these limitations, where the relatively large lattice mismatch between different materials can be relaxed due to the weak vdW forces between 2D material layers providing virtually unlimited material combinations[17–19] (vdW epitaxy).

We report here a combined synchrotron-based x-ray and angle-resolved photoemission spectroscopy (XPS/ARPES) study of the electronic properties and band structure of monolayer (ML) h-BN grown on highly oriented pyrolytic graphite (HOPG) using molecular beam epitaxy (MBE). This study was also supported by micro-Raman spectroscopy and scanning electron microscopy (SEM) for morphological and elemental assessments. Despite the presence of multiple domains (causing in-plane rotational disorder) and structural defects, electronic band dispersions were clearly observed along the high symmetry direction, reflecting the high quality of h-BN films. The valence band maximum (VBM) at the K point lies 2.8 eV below the Fermi level suggesting that the h-BN layers are undoped.

Among different substrates, the use of graphite provides several potential advantages for the growth of h-BN. Figure 1(a) presents the crystal structure of the layered h-BN. Graphite and h-BN display in-plane hexagonal symmetry with lattice constants of $a_{h-BN}$ = 0.250 nm and of $a_{Graphite}$ = 0.246 nm making them highly compatible materials. Single layer h-BN was grown on HOPG using high-temperature MBE as described previously[20,21]. The graphite substrate is a very inert surface ensuring the presence of only vdW interactions between h-BN and the substrate[22]. Figure 1(b-c) shows SEM images for a growth temperature of 1480 ℃. Bright lines (indicated by the blue arrows in figure 1(b) are observed using SEM which correspond to the positions of the graphite step edges and indicate that the h-BN primarily nucleates from these regions. The h-BN ribbons form a partial single layer along with very small multi-layer regions (bilayer or trilayer h-BN) as shown in figure 1(c). The ribbons are ~1 µm wide and ~ 50 µm long and are in agreement with our previous observations of h-BN grown on HOPG under similar conditions[21,23]. It is worthwhile noting that some of these ribbons. It is worthwhile noting that some of these ribbons are much wider than the underlying step distances on the substrate (~500 nm), suggesting that the h-BN then grows laterally across the terraces before coalescing to form a uniform layer. The lateral growth of h-BN is confirmed in Figure 1(c), which displays large h-BN monolayers, suggesting that the domains grow partially from secondary nucleation events, although we cannot exclude the formation of a domain boundary at low scale resulting from the coalescence of different h-BN domains as reported in similar studies using atomic force microscopy[20].

Using high resolution X-ray photoelectron spectroscopy (HR-XPS), we investigated the atomic composition as well as the chemical bonding environment of our samples. The XPS measurements (Figure S1) performed over a wide energy range (hv = 825 eV) showed only a small amount of oxygen contamination is present on the sample after the annealing process. The growth of the h-BN monolayer on the HOPG substrate was confirmed by the presence of the B 1s and N 1s core level peaks. High resolution spectra are shown in Figure 2(a-c). The experimental data points are displayed as dots. The solid line is the envelope of the fitted components. The underlying HOPG gives rise to a unique narrow (FWHM = 0.8 eV) carbon C1s (binding energy (BE) = 284.4 eV) peak which presents a characteristic dissymmetric shape, *i.e*, a broadening on its high energy side. This shape is expected for a conductive material and well described using a Doniach-Sunjic fitting curve considering an asymmetry value α = 0.14[24]. In addition to this main peak, an inter-band π-π* transition peak is present at higher binding energy (about 6 eV shift)[24]. Also, one main sharp peak (FWHM ~ 1.2 eV) is present on the B 1s and N 1s spectra at a BE of 190.5 eV and 398.1 eV, respectively. Overall these results suggest a main chemical environment corresponding to the B-N bond in the hexagonal structure present in an h-BN layer[25–30]. The hexagonal structure of the BN sample is confirmed by the presence in both spectra of a peak at higher binding energy (~ 9 eV) than the main peak, which corresponds to a π plasmon (in light blue color). This loss feature is present only in hexagonal BN as there are no π electrons in cubic BN[31].

The presence of grain boundaries in the sample can be related to the presence of defects and preferential sites for adsorbates (e.g. oxygen). Recent works [32,33] have shown that in the case of h-BN the grain boundaries are mostly formed by pentagon-heptagon (5-7) ring fusion which introduces homo-elemental bonding (either B-B or N-N). The presence of these two types of defects is clearly visible in the HR-XPS deconvolution of the B 1s and N 1s spectra. At first, a small shoulder located at + 0.9 eV, with respect to the main peak, is present in the B 1s spectrum. This component, representing 17% of the total photoemission intensity, can be attributed to a nitrogen vacancy which is replaced by an oxygen atom forming a B-O bond[34,22]. In the B 1s spectrum, another small component is present at lower binding energy than the main peak (-1.3 eV). This peak, as the one located at +1.4 eV in the N 1s spectrum, is the signature of a 5-7 ring attributed to the presence of grain boundaries. In particular these features indicate the formation of homo-nuclear bonds B-B and N-N, respectively[34–36]. No other components are present in the B 1s and N 1s spectra at lower binding energies, indicative of B-C[37,38] or N-C[37,38] or at higher energies, indicating N-O[22,38,39] bonds. Therefore, besides the presence of grain boundaries and nitrogen vacancies, induced naturally by the island conformation of the h-BN flakes, the sample shows high crystalline quality. In fact, the comparison of these HR-XPS spectra with our previous study carried out on a bulk h-BN flake[40] shows that the FWHM of the main XPS components are of the same order of magnitude (1 eV for the bulk h-BN and 1.2 eV in the case of polycrystalline single layer h-BN). The measurement of the secondary electron cut off (Figure 2(d)) allows the determination of the work function (ϕ) of our h-BN flakes grown on the HOPG substrate, , $\phi_{h-BN/HOPG}$ = 4.70 ± 0.05 eV. This value is really sensitive to the h-BN substrate and the type of interface bonding[41–43]

This high degree of crystalline order is also confirmed by the NEXAFS analysis (Section S3, supporting information). The angle-dependent NEXAFS spectra at the nitrogen K-edge (Figure S3(a)) and at B K edge (Figure S3(b)) show a strong dichroism of the π* and σ* resonances, clearly indicating that the h-BN layers grow parallel to the graphite substrate. In order to gain insight into the strength of the interaction between h-BN and graphite, and to determine the electronic properties of the h-BN, ARPES was used to measure the band structure of the single layer h-BN on graphite[40,44]. The ARPES intensity map was measured perpendicular to the ΓK direction of the h-BN Brillouin zone (figure 3(a)) in order to discriminate between the different h-BN and graphite domains orientations. Figure 3(b) shows the dispersion of the h-BN bands measured on the graphite substrate. The lower intensity bands between -2 eV and $E_F$ are due to the π bands of graphite domains (figure 3(c)). The figure shows two bright and one faint intersecting Dirac cones. These arise from different graphite domains within the HOPG surface; the difference in their relative orientation is reflected in the different in-plane wave-vector on which they are centered (white arrows). Note that the spot size used for the ARPES measurements is ~50 μm which is sufficient to illuminate an area on the surface containing more than one HOPG domain. Band maps on different area of the sample show similar features: multiple rotated linearly dispersing Dirac cones. The h-BN VBM overlaps with the π graphite bands, but can be seen to be ∼2.8 eV below the Fermi level consistent with a 6 eV band gap and similar to that found for single layer h-BN[6]. The high intensity and sharpness of the π bands attest the high quality of the h-BN sample. The presence of a single π band is a clear signature of single layer h-BN band structure[45]. This is confirmed by the location of this VBM at the K point, in contrast to multilayer h-BN[40]. Similar to the HOPG substrate we find domains of h-BN with different orientations suggesting the presence of rotational disorder and twinning domains. However, there is a local match of the alignment between the h-BN and the graphite; that is the valence band maximum of each h-BN domain occurs at the same in-plane vector as the Dirac point of the graphite implying that lattice vectors of the h-BN and graphite are locally orientationally aligned. Based on the difference of the wave vector between the two domains, we can estimate the rotation angle of the particular areas which contribute to the measurements in figure 3, the angle is estimated of about 10°. These results indicate that, despite a weak interaction between h-BN and graphite, van der Waals epitaxy defines the long range ordering of h-BN even on graphite. Note that AFM images published previously have identified small h-BN domains which grow with different orientational alignment[18,19]. The results present here show that when averaged over large areas orientational alignment does occur, suggesting possibly that islands with orientational alignment grow more rapidly than misaligned islands. On the basis of these ARPES and HR-XPS measurements, the band alignment of the h-BN/HOPG system (Figure 4) was also determined. This band alignment is expected to differ to the case of 3D materials junction. In fact, the low dimensionality of 1 ML h-BN (few Å thick) prohibits the formation of a depletion region. This region is replaced by what is called a van der Waals gap, as shown in Figure 4. Moreover, the lack of dangling bonds on both surfaces (HOPG and h-BN) implies the absence of surface states (no Fermi level pinning); no bonds are present at the interface between the two materials (as confirmed also by XPS measurements Figure 2) which are held together only by vdW interaction. Then, the band alignment is governed by the Schottky-Mott limit (S parameter close to 1) [46–48]. However, although vdW interaction is week, it can redistribute the charge density at the interface can induce an interface dipole [49] $\Delta V = \phi_{h-BN/HOPG} - \phi_{HOPG}$ = 4.7- $4.6^{50}$ = 0.1 eV. These results show the formation of a vdW heterostructure where the MBE grown h-BN is un-doped and electronically decoupled from the surface, maintaining its insulating character.

In summary, we have studied the electronic properties and band structure of single layer h-BN grown on graphite using MBE. We find that this heterostructure gives rise to sharp bands, in particular for the h-BN valence bands at the K points. Our ARPES measurements on the heterostructure showed that graphite and h-BN largely retained their original electronic structure. The obtained epitaxial films of single layer h-BN demonstrate very high structural perfection as confirmed using HR-XPS and NEXAFS. ARPES experiments demonstrate clearly the energy dispersion of the valence band of single layer h-BN below the Fermi level. The h-BN layer is electronically decoupled from the graphite surface, as expected in van der Waals epitaxy of a layered material such as h-BN.

**Supporting Information:**

The Supporting Information is available free of charge on the APL Publications website at DOI:
Supplementary figures: Methods and characterization, Figure S1: Raman Spectroscopy study of h-BN/graphite sample. Figure S2: Wide energy range HR-XPS spectrum of h-BN/HOPG collected at a photon energy hν = 825 eV. Figure S3: Figure S3: Partial yield absorption of a) nitrogen and b) boron K-edge NEXAFS spectra as a function of the angle between the polarization vector $\vec{E}$ and the surface normal $\vec{n}$ (θ).

**Acknowledgements:** This work was supported by the H2DH grants. We acknowledge support from GANEX (Grant No. ANR-11-LABX-0014) and Labex "Nanosaclay (Grant No. ANR-10-LABX-0035)". GANEX belongs to the public funded "Investissements d'Avenir" program managed by the French National Research Agency. This work was partly supported by the Government of the Russian Federation (contract 14.W03.31.0011 at the IOFFE Institute of RAS). This work was supported by the Engineering and Physical Sciences Research Council [Grant numbers EP/L013908/1 and EP/P019080/1]; and the Leverhulme Trust [Grant number RPG-2014-129]. PHB and SVN acknowledge the help of Dr. T.S. Cheng, Dr. Y.J. Cho, Dr. E.F. Smith and Dr. C.J. Mellor.

**Figure captions:**

**Figure 1:** a) Crystal structure of a single layer of h-BN indicating the h-BN lattice constant. b) SEM image of h-BN growth on HOPG showing nucleation from HOPG terrace steps indicated by the blue arrows. The darkest contrast corresponds to the underlying HOPG substrate and the lighter regions correspond to regions of h-BN growth. c) High resolution SEM image of an h-BN island showing regions of single and bi-layer growth and a region of exposed HOPG substrate.

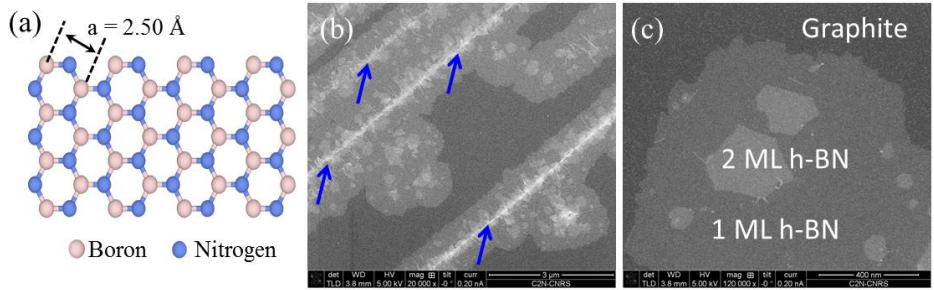

**Figure 2:** HR-XPS spectra of a) carbon 1s b) boron 1s and c) nitrogen 1s peaks (in insert a zoom of the π plasmon), respectively, measured using hv = 825 eV d) Secondary electron intensity as a function of the kinetic energy above the Fermi level for the h-BN/graphite heterostructure measured with hv = 340 eV.

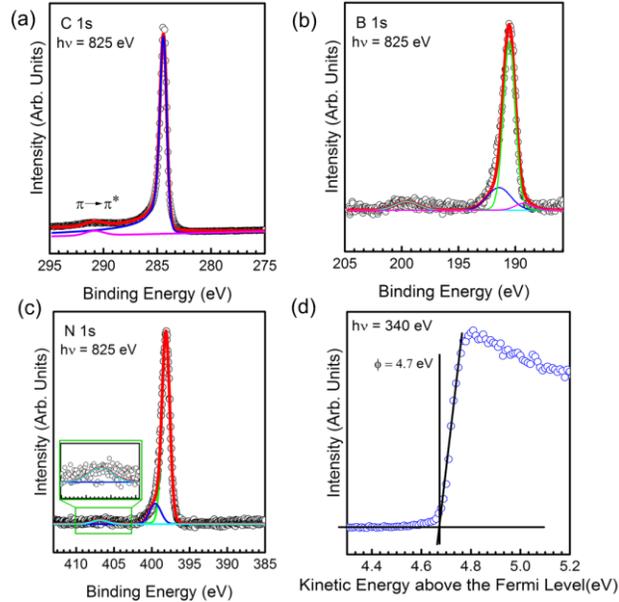

**Figure 3**: (a) h-BN Brilloion zone as seen in the ARPES map of (b) due two h-BN domains. $K_1$ and $K_2$ indicate the K point relative to these domains. The red arrow represents the measurement geometry in k-space. b) ARPES measurements of h-BN and c) Zoom of ARPES measurements in b) showing the band structure of the graphite substrate capped with h-BN at hv = 60 eV and 8 K. The white arrows point the different K points related to the different graphite domains. The Fermi level position is located at zero binding energy (indicated by the white dashed line in both b) and c)).

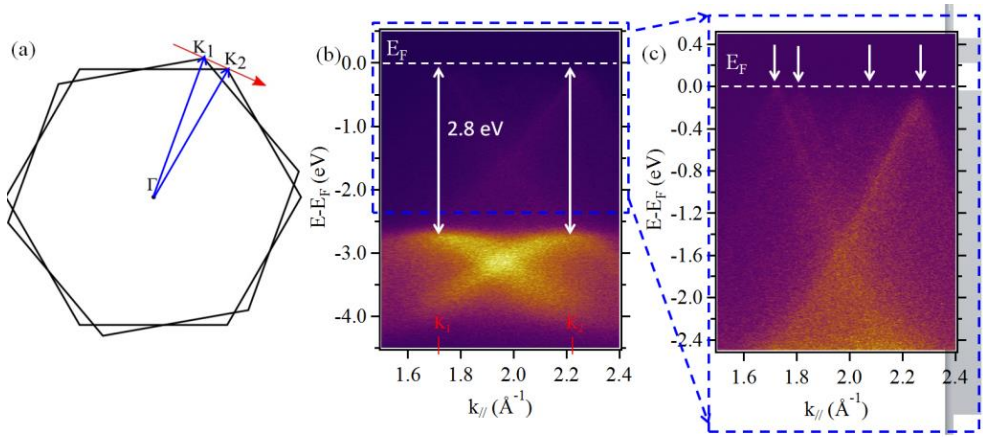

**Figure 4:** Band alignment diagram of h-BN/graphite obtained from ARPES and HR-XPS measurements, The h-BN VBM is located at -2.8 eV respect to the Fermi level and the work function of the system is 4.7 eV. A small dipole is present at the interface $\Delta V = 0.1$ eV.

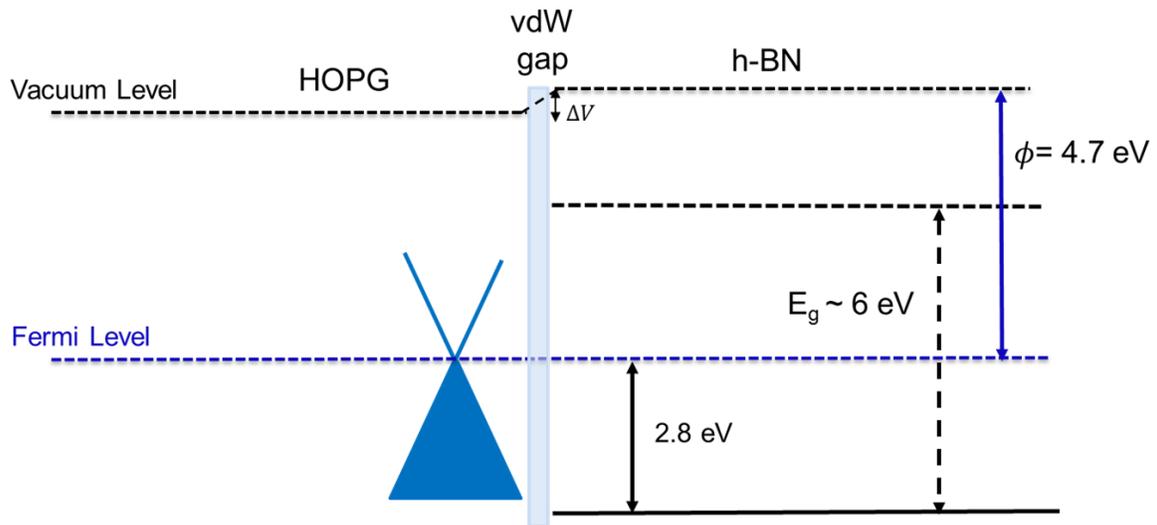

# Supporting information for "Van der Waals Epitaxy of Two-Dimensional Single-Layer h-BN on Graphite by Molecular Beam Epitaxy: Electronic Properties and Band Structure"

Debora Pierucci[1], Jihene Zribi[2], Hugo Henck[2], Julien Chaste[2], Mathieu G. Silly[3], François Bertran[3], Patrick Le Fevre[3], Bernard Gil[4], Alex Summerfield[5], Peter H. Beton[5], Sergei V. Novikov[5], Guillaume Cassabois[4], Julien E. Rault[3], and Abdelkarim Ouerghi[2]

[1]CELLS - ALBA Synchrotron Radiation Facility, Carrer de la Llum 2-26, 08290 Cerdanyola del Valles, Barcelona, Spain
[2]Centre de Nanosciences et de Nanotechnologies, CNRS, Univ. Paris-Sud, Université Paris-Saclay, C2N – Marcoussis, 91460 Marcoussis, France
[3]Synchrotron-SOLEIL, Saint-Aubin, BP48, F91192 Gif sur Yvette Cedex, France
[4] Laboratoire Charles Coulomb UMR 5221, Université de Montpellier & CNRS, F-34095, Montpellier, France
[5]School of Physics and Astronomy, University of Nottingham, Nottingham NG7 2RD, United Kingdom

**Methods and characterization:**

Sample preparation:

All h-BN layers investigated in this paper were grown by MBE using a high-temperature effusion cell for sublimation of boron and a standard radio-frequency (RF) plasma source for active nitrogen flux[1,2]. We performed MBE growth for 3 hours on 10 × 10 mm² HOPG substrates with a mosaic spread of 0.4° at a substrate temperature of 1480°C. Details of the high-temperature MBE system are described elsewhere[1,2].

Micro-Raman spectroscopy:

Micro-Raman measurements were carried out using a commercial confocal Renishaw with a 100× objective, a Si detector (detection range up to ~2.2 eV) and a 532 nm laser at room temperature. In addition, the excitation laser was focused onto the samples with a spot diameter of ~1 μm.

Photoemission spectroscopy:

XPS/NEXAFS experiments were performed in ultra-high vacuum at the SOLEIL Synchrotron facility (Saint-Aubin, France) on the Tempo beamline. During the XPS measurements, the photoelectrons were detected at 0° from the sample surface normal $\vec{n}$ and at 46° from the polarization vector $\vec{E}$. A Shirley background was subtracted in all core level spectra. The B 1s and N 1s spectra were fitted using sums of Voigt curves composed by the convolution of Gaussian and Lorentzian line shapes[3]. The Lorentzian full-width at half-maximum was fixed at 70 meV for B 1s and at 115 meV for N 1s. NEXAFS measurements were performed in total electron yield. The spectra were measured for different incidence angles of the linear polarized synchrotron light. We indicate with θ the angle between the polarization vector $\vec{E}$ and the surface normal $\vec{n}$, (θ = 10°, 45° and 90°). ARPES measurements were performed at the Cassiopée beamline at 8 K, the photon energy (hν = 60 eV) and sample orientation were set in order to explore the k-space region around the K point in the perpendicular ΓK direction of the graphite Brillouin zone.

### S1. Raman Spectroscopy study:

Figure S1 shows a typical micro-Raman spectrum of our sample. The contributions of graphite are identified by three main structures: i) the D band at 1370 cm$^{-1}$ (the small D peak amplitude indicates the high quality of the sample[4]), ii) the G band at 1595 cm$^{-1}$ and iii) the 2D band at 2720 cm$^{-1}$. The D band shows an asymmetric shape (inset of figure 1(d)), indicating the presence of another component at ∼1367 cm$^{-1}$ (Lorentzian fit). This feature is the signature of the $E_{2g}$ mode of h-BN vibrations[5–7] and occurs at the same value, ~1367 cm$^{-1}$, as expected for an h-BN crystal[8], suggesting that no stress was induced by the graphite substrate. The full width at half maximum (FWHM) of this peak, estimated from a Lorentzian fitting, is 12 ± 0.5 cm$^{-1}$. The small linewidth of the narrow Raman peak indicates that the h-BN layer has high crystalline quality[9]. Moreover, only this $E_{2g}$ mode was observed confirming that no phases other than h-BN are present (note that cubic BN would give rise to two other Raman modes at 1056 cm$^{-1}$ (TO phonon) and 1306 cm$^{-1}$ (LO phonon)). The relative small peak amplitude, compared to the substrate, indicates a very thin film as expected for single layer h-BN. Thus, our measurements show that the quality of our h-BN films is similar to that of single crystal bulk h-BN[10].

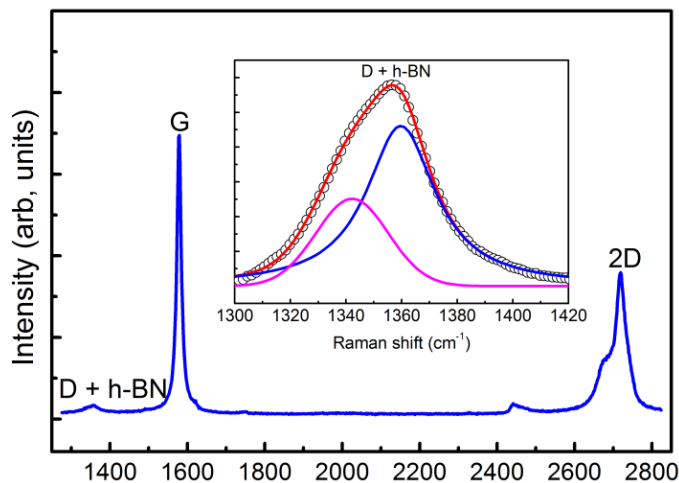

**Figure S1:** Raman spectrum taken on h-BN/HOPG. (inset) Zoom of Raman spectra showing the contribution of D mode of graphite and $E_{2g}$ mode of h-BN.

### S2. XPS Overview:

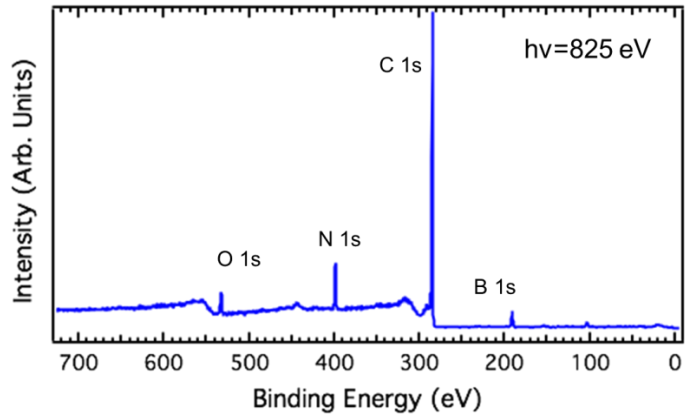

**Figure S2:** Wide energy range HR-XPS spectrum of h-BN/HOPG collected at a photon energy hν = 825 eV.

## S2. NEXAFS spectra:

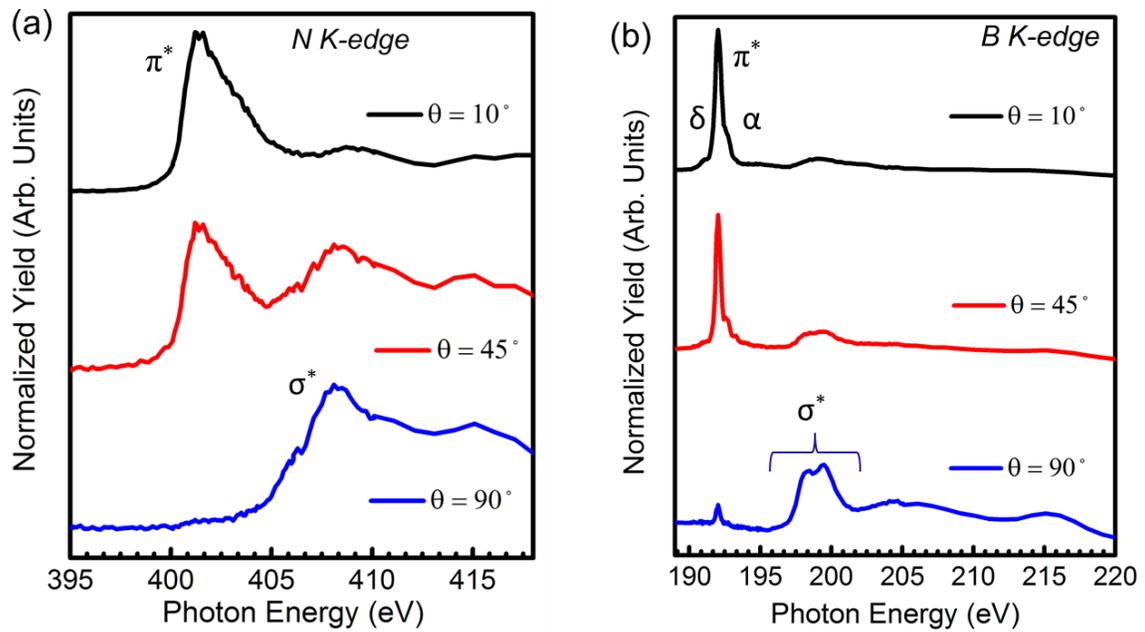

**Figure S3**: Partial yield absorption of a) nitrogen and b) boron K-edge NEXAFS spectra as a function of the angle between the polarization vector $\vec{E}$ and the surface normal $\vec{n}$ (θ).

We define θ, the angle between the polarization vector $\vec{E}$ and the surface normal $\vec{n}$ (θ = 10°, 45°, and 90°). For the NEXAFS around the K-edge, the initial state involved in the transition is a 1s state, so the final state must contain contributions from a *p* orbital, for instance in the form of $s+p_z$ (named π* transitions) or $p_x+p_y$ (σ* transitions) due to the dipole selection rule Δl = ±1 (where Δl is the change of the angular momentum). Since the π* and σ* final states are normal to each other, we expect that in the case of perfect flat h-BN flakes, these two components evolve in an opposite trend as θ varies (i.e. varying the orientation of $\vec{E}$). The nitrogen K-edge spectra are shown in figure S3(a). An excitonic N ($1s \rightarrow \pi^*$) peak with a maximum centred at ~401.4 eV and three $\sigma^*$ transitions located at ~405.8 eV, ~408.1 eV and 415 eV are observed in these spectra, as reported in the literature for bulk h-BN[11,12]. Figure S3(b) shows the boron K-edge spectra. Additionally to the characteristic B ($1s \rightarrow \pi^*$) excitonic transition of h-BN at ~192.0 eV (which is a specific fingerprint of $sp^2$-hybridized B atoms in the hexagonal h-BN network, *i.e.*, a trigonal B-$N_3$ bonding environment) and the two $\sigma^*$ peaks at ~198.2 eV and ~199.5 eV, another excitonic $\pi^*$ peak is observed at ~192.7 eV. This peak labelled α is assigned to a B atom surrounded by 1 nitrogen vacancy, filled by oxygen, corresponding to the B-$N_2$-O configuration[13]. This observation is in agreement with, and reinforces greatly the HR-XPS results. Moreover, in the spectrum we clearly see a peak, which is located below the adsorption edge of h-BN, i.e. within the energy gap. This component, labelled δ, is located at ~191 eV. Orellana et al.[14] predicted, in the case of B-B bonds, the presence of several occupied, half-occupied and empty levels inside the band gap of h-BN related to this defect. In particular, the empty level should be located at +2.98 eV with respect to the valence band maximum (VBM). In our h-BN sample, the VBM is located at about 187.6 eV (the VBM obtained by ARPES measurement, see figure 3(a), is here referred to the ionization potential IP = 195.2 eV) thus positioning this empty level at about 190.6 eV. This theoretical prediction is very close to the position of our δ peak.